\documentclass[12pt]{article}
\usepackage{amssymb,amsmath,cite}
\long\def\@makefntext#1{\parindent 0cm\noindent
\hbox to 1em{\hss$^{\@thefnmark}$}#1}

\begin{document}
\begin{titlepage}
\vspace{.5in}
\begin{flushright}
\end{flushright}
\vspace{.5in}
\begin{center}
{\Large\bf
 Exact dark energy star solutions }\\
\vspace{.4in}
{Stoytcho~ S.~Yazadjiev\footnote{\it email: yazad@phys.uni-sofia.bg}\\
       {\small\it Department of Theoretical Physics, Faculty of Physics,}\\
       {\small\it Sofia University, Sofia, 1164, Bulgaria }\\}
\end{center}

\vspace{.5in}
\begin{center}
{\large\bf Abstract}
\end{center}
\begin{center}
\begin{minipage}{4.7in}
{\small Adopting  the phantom (ghost) scalar field description of dark energy,  we construct a general class of exact interior solutions describing mixed relativistic
stars containing both ordinary matter and dark energy in different proportions. The exterior solution that continuously matches the interior solutions is also found.
Exact  solutions describing  extremal configurations with zero ordinary matter pressure are also constructed. } \\ \\
 PACS:  04.40.Dg;  95.36.+x; 04.20.Jb
\end{minipage}
\end{center}
\end{titlepage}
\addtocounter{footnote}{-1}

The astronomical  observations of the present Universe provide  evidence for the existence of
a mysterious kind of matter called dark energy which governs the expansion of the Universe \cite{R1},\cite{ASSS}.
The dark energy exhibits some unusual properties such as negative pressure to density ratio $w$ (in hydrodynamical language) and violation of the energy conditions.
The ratio $w$ may be less than $-1/3$ corresponding to violation of the strong energy condition, or even less
than $-1$ which is violation of  the weak  or null energy condition.  Current experimental data shows that $w$ is in the range  $-1.38< w < -0.82$ .

The fundamental role that the dark energy plays in  cosmology naturally makes us  search for local astrophysical manifestation of it.
In the present work we consider models of relativistic stars containing not only ordinary matter but also dark energy.  The existence of dark energy
makes us  expect that the present-day existing stars are a mixture of both ordinary matter and dark energy  in different proportions.
The study of such mixed objects is a new interesting problem and some steps in this direction have already been made (see for example  \cite{MM}-\cite{DFKK} and references therein).

A possible theoretical description of the dark energy is provided by scalar fields with negative kinetic
energy, the so-called phantom (ghost) scalars \cite{Caldwell},\cite{CSSX}. The negative kinetic energy, however,
leads to severe quantum instabilities\footnote{The perfect fluid description of the dark energy also suffers from instabilities due to
the imaginary velocity of the sound. From a classical point of view the massless phantom field is even more stable than  its usual counterpart \cite{BCCF},\cite{APicon}.  }
and this is a formidable challenge to the theory. However, there are claims that these instabilities can be avoided \cite{PT}.  In general, the problem could be avoided if we consider the phantom scalars as an effective field theory resulting from  some kind of fundamental theory with a positive energy \cite{NO},\cite{CHT}. In this case the phantom scalar description of the dark energy is physically acceptable. In this context it is worth noting that the phantom-type fields arise in string theories and  supergravity \cite{Sen1}-\cite{Nilles}.

In this work we adopt a description of the dark energy by a phantom scalar. Then the Einstein equations in the presence of dark energy read
\begin{eqnarray}\label{FE}
&&R_{\mu\nu}= 8\pi (T_{\mu\nu} - {1\over 2}Tg_{\mu\nu}) - 2\partial_{\mu}\varphi\partial_{\nu}\varphi,\\
&&\nabla_{\mu}\nabla^{\mu}\varphi= 4\pi \rho_{D}. \nonumber
\end{eqnarray}
Here $T_{\mu\nu}$ is the energy-momentum tensor of the ordinary matter in the perfect fluid description with energy density $\rho$ and pressure $p$:
\begin{eqnarray}
T_{\mu\nu}=(\rho + p)u_{\mu}u_{\nu} + pg_{\mu\nu}.
\end{eqnarray}
For the ordinary matter we impose the  natural conditions $\rho\ge 0$ and $p\ge 0$.   The density of the dark energy sources is denoted by $\rho_{D}$.
The dark energy sources will be called {\it dark charges}.

Since a very little is kown about the interaction of dark energy with the normal matter, in our model we consider only minimal interactions for the phantom field --
 we have not included terms describing non-minimal  interaction
between the phantom field and the normal matter in the field equations (\ref{FE}).

When we consider the local manifestation of  dark energy on astrophysical scales, i.e. scales much smaller than the cosmological scales,
the phantom potential ${\cal U}(\varphi)$ can be neglected and that is why we  set ${\cal U}(\varphi)=0$. This is not a principle restriction.
Exact solutions with the desired properties  as  those presented below can be also found in the  presence of phantom potential and non-minimal interaction
between the ordinary matter and the phantom scalar. Since we  do not seek mathematical generality but we are interested in the physics of the model we shall
restrict ourselves to the case ${\cal U}(\varphi)=0$ and minimal interaction with the ordinary matter.

In what follows  we give  exact solutions describing  mixed relativistic  dark energy stars.
The dark star solutions are characterized by the mass $M$, the dark charge $D$ and the (coordinate) radius $R$.

We consider static, spherically symmetric and asymptotically flat space-time with a metric
\begin{eqnarray}\label{metric}
ds^2= - e^{2U}dt^2 + e^{-2U+2\lambda} \left[e^{-2\chi}dr^2 + r^2(d\theta^2 + \sin^2\theta d\phi^2) \right],
\end{eqnarray}
where the metric functions are functions of the radial coordinate only. The phantom field $\varphi$ and the dark charge  density $\rho_{D}$ are required to be
static and to depend on the radial coordinate $r$ only.   For the normal matter  fluid we impose the usual  conditions for staticity and spherical symmetry
 $p=p(r)$, $\rho=\rho(r)$ and $u_{\mu}dx^{\mu}=-e^{U}dt$.
Applying the mathematical techniques developed in \cite{Y} we can generate exact interior  solutions to the field equations (\ref{FE}) using  any exact
interior solution of the ordinary Einstein-perfect-fluid equations as a seed. More precisely, we have the following proposition:

\medskip
\noindent

{\bf Proposition}\;
{\it Let}
\begin{eqnarray}\label{SEEDSOL}
&&ds^2_E= - e^{2\lambda}dt^2 + e^{-2\chi}dr^2 +  r^2(d\theta^2 + \sin^2\theta d\phi^2), \nonumber \\
&&p_{E}=p_{E}(r), \\
&&\rho_{E}=\rho_{E}(r) \nonumber
\end{eqnarray}
{\it be an interior solution to  the ordinary Einstein-perfect-fluid equations, then the metric (\ref{metric}) together with the functions}

\begin{eqnarray}\label{DSSOL}
&&e^{2U}= e^{2\lambda\cosh\beta}, \nonumber \\
&&\rho= e^{2U-2\lambda} \left[\rho_E \cosh\beta + 3(\cosh\beta-1)p_E\right], \nonumber\\
&&p=e^{2U-2\lambda}p_E, \\
&&\rho_{D}= e^{2U-2\lambda} \left(\rho_{E} + 3p_E\right)\sinh\beta,  \nonumber \\
&&\varphi=\sinh\beta \lambda \nonumber
\end{eqnarray}
{\it form an interior solution to the field equations (\ref{FE}) where $\beta$ is an arbitrary real constant}.

\medskip
\noindent
It is worth noting that  the solution generating method described in the proposition can be applied for an arbitrary equation of state of the ordinary matter.

If the seed solution (\ref{SEEDSOL}) has a well-defined boundary $r=R$, where by definition $p_E(R)=0$, the same is true
for the solution (\ref{DSSOL}), namely $p(R)=0$. Therefore $R$ can be interpreted as the coordinate radius of the dark energy star.
On the dark star surface $r=R$, our interior solution (\ref{DSSOL}) matches continuously the following exterior solution

\begin{eqnarray}\label{external}
&&ds_{ext}^2= - \left(1- {2m\over r}\right)^{\cosh\beta} dt^2
+ \left(1- {2m\over r}\right)^{1-\cosh\beta}\left[ \frac{dr^2}{1-\frac{2m}{r}} + r^2(d\theta^2 + \sin^2\theta d\phi^2)\right], \nonumber\\
&&\varphi_{ext}= {1\over 2}\sinh\beta \ln\left(1-\frac{2m}{r} \right).
\end{eqnarray}
For $\beta=0$ this solution reduces to the vacuum Schwarzschild solution  describing the  exterior gravitational field of a static and spherically symmetric star with mass $m$.
The exterior solution (\ref{external}) can be easily and elegantly generated from the  vacuum Schwarzschild solution if the $SO(1,1)$ symmetry of the dimensionally reduced
field equations (\ref{FE})  with $\rho=p=\rho_D=0$ is used. However, it should be mentioned that (\ref{external})  was obtained for the first time in \cite{BL} (with another method)
and can be considered as a phantom counterpart of Fisher's scalar vacuum solution for a normal scalar \cite{F},\cite{B}.

The mass and the dark charge of the dark star are given by

\begin{eqnarray}
M&=& -{1\over 4\pi}\int_{Star} R_{t}^{t}\sqrt{-g}d^3x =\int_{Star} (\rho + 3p)\sqrt{-g}d^3x =  \\
&& \cosh\beta\int_{Star} e^{2\lambda}(\rho_{E} + 3p_{E})\sqrt{-g_{E}}d^3x=\cosh\beta \,m , \nonumber\\ \nonumber \\
D&=&\int_{Star} \rho_{D}\sqrt{-g}d^3x={1\over 4\pi} \oint_{S_{\infty}^2} \nabla_{\mu}\varphi d\Sigma^{\mu}= \\
&&\sinh\beta \int_{Star} e^{2\lambda}(\rho_{E} + 3p_{E})\sqrt{-g_{E}}d^3x =\sinh\beta \, m \nonumber
\end{eqnarray}
The same expressions for the mass and the dark charge are also obtained from the asymptotic expansion of the exterior solution.
The mass $M$ and the dark charge $D$  satisfy the inequality $M>|D|$ as one can see from the above expressions.
The dark charge $D$ is a measure of the content of dark energy in the star. When there is no dark energy in the star (i.e. $D=0$)
our solution reduces to the seed interior  solution describing  ordinary perfect fluid star in general relativity.
The initial solution parameters $m$ and $\beta$ can be expressed in terms of  $M$ and $D$ as follows

\begin{eqnarray}
m=\sqrt{M^2-D^2}, \;\;\;
\cosh\beta= \frac{M}{\sqrt{M^2-D^2}}.
\end{eqnarray}

In the case of ordinary stars ($D=0$), the radius of the star $R$ and the gravitational  Schwarzschild radius $R_G=2m$ satisfy the  well-known Buchdahl inequality \cite{Buchdahl}

\begin{eqnarray}\label{RatioGR}
\frac{2m}{R}<{8\over 9}.
\end{eqnarray}

For the mixed dark stars we have

\begin{eqnarray}\label{MRRatio}
\frac{2M}{R_{ph}}= \frac{2m\cosh\beta}{R}\left(1- \frac{2m}{R}\right)^{{(\cosh\beta -1)/2}},
\end{eqnarray}
where $R_{ph}= R\left(1- \frac{2m}{R}\right)^{(1-\cosh\beta )/2}$ is the physical radius of the dark star. By examining the right hand side of  eq.(\ref{MRRatio})
and taking into account eq.(\ref{RatioGR}) one can show that for arbitrary $\beta$ we have

\begin{eqnarray}
\frac{2M}{R_{ph}} <{8\over 9}.
\end{eqnarray}

 As an illustration of the above presented method for generating exact dark energy star solutions we will give
 a fully explicit dark energy star interior solution generated from the well-known interior Schwarzschild solution \cite{KSHM}.
 The interior Schwarzschild solution qualitatively describes the general
case of a static, spherically symmetric perfect fluid star in
general relativity and, therefore we expect that the generated solution should qualitatively describe  the general case of mixed dark stars.
With the generation techniques applied on the   interior  Schwarzschild solution we obtain:

\begin{eqnarray}\label{SINT}
&&ds_{int}^2 = - e^{{2M\lambda \over\sqrt{M^2-D^2}}} dt^2 +
e^{{-2{M-\sqrt{M^2-D^2}\over \sqrt{M^2-D^2} }}\lambda }\left[\frac{dr^2}{1 - \frac{2\sqrt{M^2-D^2}}{R^3}r^2} +
r^2 \left(d\theta^2 + \sin^2\theta d\phi^2 \right) \right],\\
\nonumber\\
&&p = {3\sqrt{M^2-D^2}\over 4\pi R^3}
e^{2{M-\sqrt{M^2-D^2}\over \sqrt{M^2-D^2}}\lambda}\left[\left(1- \frac{2\sqrt{M^2-D^2}}{R^3}r^2\right)^{1/2} -
\left(1-  \frac{2\sqrt{M^2-D^2}}{R} \right)^{1/2}\over 3\left(1-  \frac{2\sqrt{M^2-D^2}}{R} \right)^{1/2} -
\left(1- \frac{2\sqrt{M^2-D^2}}{R^3}r^2\right)^{1/2}\right],\\
\nonumber\\
&&\rho= {3M\over 4\pi R^3 } e^{2{M-\sqrt{M^2-D^2}\over \sqrt{M^2-D^2}}\lambda }  + 3{M-\sqrt{M^2-D^2}\over \sqrt{M^2-D^2}}p ,\\
\nonumber\\
&&\rho_{D} = {3D\over 4\pi R^3}e^{2{M-\sqrt{M^2-D^2}\over \sqrt{M^2-D^2}}\lambda } +  {3D\over \sqrt{M^2-D^2}}p,\\
\nonumber\\
&&\varphi= {D \over \sqrt{M^2-D^2}}\, \lambda,
\end{eqnarray}
where
\begin{eqnarray}
e^{\lambda}= \left[{3\over 2}\left( 1- {2\sqrt{M^2-D^2}\over R} \right)^{1/2}
- {1\over 2}\left(1 - {2\sqrt{M^2-D^2}\over R^3}r^2\right)^{1/2} \right].
\end{eqnarray}

Contrary to the pressure, the fluid energy density $\rho$ and the dark charge density $\rho_{D}$ do not vanish on
the boundary just as in the case of the fluid energy density in the interior Schwarzschild solution. We have
\begin{eqnarray}
&&\rho(R) = {3M\over 4\pi R^3 } \left(1- {2\sqrt{M^2-D^2}\over R} \right)^{ {M-\sqrt{M^2-D^2}\over \sqrt{M^2-D^2}}},\\
&&\rho_{D}(R)= {3D\over 4\pi R^3 } \left(1- {2\sqrt{M^2-D^2}\over R} \right)^{ {M-\sqrt{M^2-D^2}\over \sqrt{M^2-D^2}}}.
\end{eqnarray}

Our interior solution is completely regular everywhere for $0\le r\le R$ if the dark star radius $R$ satisfies the inequality
\begin{eqnarray}\label{YIEQ}
{\sqrt{M^2-D^2}\over R} < {4\over 9}.
\end{eqnarray}

On the dark star surface $r=R$, our interior solution (\ref{SINT}) matches continuously the  exterior solution (\ref{external}) which in terms of $M$ and $D$ reads
\begin{eqnarray}\label{EXSIN}
ds_{ext}^2 &=& - \left(1- {2\sqrt{M^2-D^2}\over r}\right)^{M\over \sqrt{M^2-D^2}}dt^2  \nonumber \\
&&+ \left(1- {2\sqrt{M^2-D^2}\over r}\right)^{-{M-\sqrt{M^2-D^2}\over \sqrt{M^2-D^2}}} \left[{dr^2\over 1 - {2\sqrt{M^2-D^2} \over r} } +
r^2(d\theta^2 + \sin^2\theta d\phi^2) \right],\\
\nonumber\\
\varphi_{ext}&=& {D\over  2\sqrt{M^2-D^2}}\ln\left(1 - {2\sqrt{M^2-D^2}\over r} \right).
\end{eqnarray}

This solution reduces to the exterior Schwarzschild solutions in the absence of dark energy (i.e. for $D=0$).

We will also consider  extremal dark star configurations which can be obtained from  (\ref{SINT}) by taking the limit $|D|\to M$ and keeping $R$ fixed. The described limit gives
the following interior solution
\begin{eqnarray}\label{EINT}
&&ds^2_{e}=- e^{- \frac{M}{R}\left(3 - \frac{r^2}{R^2}\right)} dt^2 + e^{\frac{M}{R}\left(3 - \frac{r^2}{R^2}\right)}\left[dr^2 + r^2(d\theta^2 + \sin^2\theta d\phi^2 ) \right],   \nonumber \\
&&\rho= \frac{M}{\frac{4\pi R^3}{3}} e^{- \frac{M}{R}\left(3 - \frac{r^2}{R^2}\right)},\nonumber \\
&& \rho_{D}= \pm \rho, \\
&&p=0,    \nonumber \\
&&\varphi= \mp \frac{M}{2R}\left( 3 - \frac{r^2}{R^2}\right). \nonumber
\end{eqnarray}
This extremal interior solution matches  continuously the  exterior solution which is obtained from (\ref{EXSIN}) in the limit $|D|\to M$, namely

\begin{eqnarray}\label{EXES}
&&ds^2_{e}= - e^{-\frac{2M}{r}}dt^2 + e^{\frac{2M}{r}}\left[dr^2 + r^2(d\theta^2 + \sin^2\theta d\phi^2 ) \right], \\
&& \varphi = \mp\frac{M}{r}.
\end{eqnarray}

As one can see from (\ref{EINT}) the pressure $p$ of the ordinary matter   vanishes in the extremal limit.
How  can one explain the existence of such extremal dark stars with zero pressure?  Generally speaking, the phantom field yields repulsive rather than attractive force.
From the contracted Bianchi identities
\begin{eqnarray}\label{Binachi}
\partial_{r}p + (\rho + p)\partial_{r}U = \rho_{D}\partial_{r}\varphi ,
\end{eqnarray}
describing the  equilibrium, one can see that the dark energy term $\rho_{D}\partial_{r}\varphi$ provides additional effective pressure which can balance the gravitational forces.

The extremal configurations in the general case can be investigated on the basis of the field equations (\ref{FE}). Following \cite{Y}
 one can show  that for extremal configurations with zero pressure ($p=0$)
 the 3-metric $h_{ij}=e^{-2U}g_{ij} (i,j=1,2,3)$ is flat (i.e $h_{ij}=\delta_{ij}$), $\rho_D=\pm \rho$, $\varphi=\pm U$ and that $U$ satisfies the equation
\begin{eqnarray}
\Delta_{f}U= 4\pi \rho e^{-2U} ,
\end{eqnarray}
where $\Delta_{f}$ is the ordinary flat Laplacian.
Solving this equation in vacuum, i.e. for $\rho=0$, and taking into account that $g_{ij}=\delta_{ij} e^{2U}$ and $\varphi=\pm U$ we obtain  the exterior solution (\ref{EXES}).
Therefore the exterior extremal solution (\ref{EXES}) is the same for all extremal configurations with zero ordinary matter pressure .

Exact  extremal interior solutions can be found by specifying the dependance $\rho=\rho(U)$. For example the extremal interior solution (\ref{EINT})  is obtained for $\rho e^{-2U}=const$. Another exact and physically well behaved solution is found for $\rho e^{-2U}= -C^2/4\pi \times U$ where $C>0$ is a constant. In this case we have to
solve the equation
\begin{eqnarray}
\Delta_{f}U + C^2 U=0.
\end{eqnarray}
The solution of this equations which is well behaved at the center $r=0$ is
\begin{eqnarray}
U= A \frac{\sin{Cr}}{r},
\end{eqnarray}
with $A$ being an integration constant. In order to match the exterior and interior solution we require that the potential $U$ and its   radial derivative are continuous on the boundary $r = R$.  Imposing these conditions we obtain the following solution

\begin{eqnarray}
U= -M \frac{\sin\left(\frac{\pi}{2}\frac{r}{R}\right)}{r}.
\end{eqnarray}
Choosing appropriate functions $\rho=\rho(U)$ one can construct many other extremal solutions.

Finally, we give the upper bound of the ratio $2M/R_{ph}$ for the extremal configurations, namely

\begin{eqnarray}
\frac{2M}{R_{ph}}= \frac{2M}{R} e^{-M/R} \le \frac{2}{e} .
\end{eqnarray}

In conclusion, the  exact interior solutions (and the exterior solution) found in the present paper could be used  in studying, both qualitatively and quantatively,
the local astrophysical effects related to the existence of dark energy. Some implications of the solutions presented in the present paper will be considered in a future publication.
There remain some issues that need further investigation. One such issue is the stability of the  found solutions. The fact that the phantom scalar field exhibits more stable
behavior than its canonical counterpart \cite{BCCF},\cite{APicon} and the results in \cite{DFKK} show that we should expect that the non-extremal solutions are classically stable
when the seed Einstein-perfect-fluid solution is stable.
This problem will be considered more thoroughly in a future publication.

The present work could also be extended in the following direcrion. Performing an approprate analytical continuatioin of the exterior solution (\ref{external})
 we obtain the well-known Bronnikov--Ellis phantom scalar (vacuum) wormhole  solution \cite{B2},\cite{E}. It is interesting whether Bronnikov--Ellis branch can also  be
the external field of a star and  what is the structure of such a star.  We anticipate that some interior solutions countiniously matching the exterior
Bronnikov--Ellis solution could be obtained by an approprate analytical continuatioin of some of the interior solutions constructed in this paper and probbaly
some of those solutions would have structure similar to that considered in   \cite{DFKK}.

\vspace{1.5ex}
\begin{flushleft}
\large\bf Acknowledgments
\end{flushleft}

This work was supported in part by  the Bulgarian National Science Fund under Grants DO 02-257, VUF-201/06 and by Sofia University Research Fund under
Grant 88/2011.

\end{document}